\documentclass{acm_proc_article-sp}

\usepackage{makeidx}  
\usepackage{latexsym}
\usepackage{amsmath}
\usepackage{amssymb}
\usepackage{algorithm,algorithmic}
\usepackage{pstricks,nicefrac}
\usepackage{epsfig}
\usepackage{graphics}
\usepackage{url}
\usepackage{pst-tree}
\usepackage{color,amsmath}
\usepackage{multirow}
\usepackage{appendix}
\usepackage{mdwlist}
\usepackage{times}

\newcommand{\ca}[1]{\mathcal{#1}}
\newcommand{\tuple}[1]{\langle{#1}\rangle}
\newcommand{\toll}{{\it toll}}

\newcommand{\ie}{i.e.,}

\newcommand{\real}[1]{\overline{#1}}
\newcommand{\claimed}[1]{{#1}}

\newcommand{\cost}{\mathit{cost}}
\newcommand{\paid}{\mathit{pay}}

\newcommand{\client}[1]{{#1}_c}

\newcommand{\hencr}[2]{{\cal E}_{#2}(#1)}

\newtheorem{definition}{Definition}
\newtheorem{theorem}{Theorem}

\newtheorem{lem}{Lemma}

\def \sharedaffiliation{%
\end{tabular}
\begin{tabular}{ccc}}
\numberofauthors{1}

\author{
\alignauthor
\begin{tabular}{cccc}
Xihui Chen,$^{\dagger,\sharp}$&
Gabriele Lenzini,$^{\dagger}$ &  
Sjouke Mauw,$^{\dagger,\ddagger}$&
Jun Pang$^{\ddagger}$
\end{tabular}
\sharedaffiliation
\affaddr{$^\dagger$Interdisciplinary Centre for Security}& 
\affaddr{$^\sharp$itrust consulting s.\`{a} r.l.,}&
\affaddr{$^\ddagger$Computer Science and}\\
\affaddr{Reliability and Trust,}& 
\affaddr{Luxembourg}&
\affaddr{Communications,}\\
\affaddr{University of Luxembourg, Luxembourg}&&
\affaddr{University of Luxembourg, Luxembourg}\\
}
\begin{document}
\conferenceinfo{SAC'12}{March 25-29, 2012, Riva del Garda, Italy.}
\CopyrightYear{2011} 
\crdata{978-1-4503-0857-1/12/03}  

\title{A Group Signature Based Electronic Toll Pricing System}

%

\maketitle

\begin{abstract}
With the prevalence of GNSS technologies, nowadays freely available
for everyone, location based vehicle services such as electronic
tolling pricing systems and pay-as-you-drive services are rapidly
growing. Because these systems collect and process travel records,
if not carefully designed, they can threat users' location privacy. 
Finding a secure and privacy-friendly solution 
is a challenge for system designers. 
In addition to location privacy,
communication and computation overhead should be taken into account
as well in order to make such systems widely adopted in practice.
In this paper, we propose a new electronic toll pricing system based on group signatures.
Our system preserves anonymity of users within groups,
in addition to correctness and accountability.
It also achieves a balance
between privacy and overhead imposed upon user devices.
\end{abstract}





\section{Introduction}
\label{Sec:intro}

Electronic Toll Pricing (ETP) systems, by collecting tolls
electronically, aim to eliminate delays due to queuing on toll roads
and thus to increase the throughput of transportation networks. Since
Norway built the first working ETP system in 1986, ETP systems have
been implemented worldwide.
Nowadays, by exploiting the availability of free Global Navigation
Satellite Systems (GNSS), traditional ETP systems are
evolving into more sophisticated location based vehicular services.
They can offer smart pricing, e.g., by charging less who drives on 
roads without congestions or during off-peak hours. 
Insurance companies can also bind insurance premiums to roads that their 
users actually use, and offer a service known as ``Pay-As-You-Drive''
(PAYD)~\cite{TDK07}.
Moreover, the collected traffic usage records can be used for public 
interest, such as planning roads maintenance, or for resolving legal disputes 
in case of accidents. 
As location is usually considered as a sensitive and private piece of information, 
ETP and PAYD systems raise obvious privacy concerns.
In addition, by processing locations and travel records, they can learn and 
reveal user sensitive information such as home addresses
and medical informations~\cite{MKW10}, which consequently can lead to material 
loss or even bodily harm.
Building secure ETP and PAYD systems that guarantee \emph{location privacy} and 
high quality of service is actually a scientific challenge.

In the last few years, secure ETP and PAYD systems have been widely
studied~\cite{TDK07,JJ08,PBB09,BRT10,GVJ11,MMCS11}.  They can roughly be
divided into two categories based on whether locations are stored in
user devices or collected by central toll servers. PriPAYD~\cite{TDK07},
PrETP~\cite{BRT10}, the cell-based solution described
in~\cite{GVJ11}, and Milo~\cite{MMCS11} belong to the first category. In these systems,
locations and tolls are managed by user devices while servers are
allowed to process only aggregated data. In the second category we
find VPriv~\cite{PBB09}, where the server stores a database of users'
travel history, and the ETP system described in~\cite{JJ08}, where 
the server collects hash values of the trip records.

Both categories have advantages and disadvantages.  Hiding
locations from servers drastically reduces concerns about location
privacy. However, the load for user devices is considerable.
Typically, devices have to manage the storage of locations and proofs to 
convince servers not to have cheated, e.g., making use of zero-knowledge 
proofs. 
On the other side, the availability of locations databases collected by servers 
like in VPriv can help improve applications such as traffic monitoring and 
control although the integration of multiple systems should be carried out 
carefully.  Whereas, solutions to preserve location privacy become a mandatory 
requirement. 
In this paper, we follow the design principles of VPriv~\cite{PBB09}.

In VPriv, users select a set of random tags beforehand and send their locations 
attached with these tags to the toll server. The server then computes and 
returns \emph{all} location fees. Each user adds up his location fees 
according to his tags and proves the summation's correctness to the server by 
using zero-knowledge proof, without revealing the ownership of the tags. 
This process needs to run several rounds every time to avoid user
behaviours deviating from the system.
Thus the main disadvantage with VPriv is that the computation and communication overhead
increases linearly with the number of \emph{rounds} executed and with
the number of \emph{users}.

\vspace{2mm}
\noindent
{\bf Our contributions.}
We propose a novel but simple ETP system which achieves a balance between 
privacy and overhead for users.  
By dividing users into groups and calculating
toll in one round, we reduce the amount of exchanged information as well as the
computation overhead due to the smaller number of locations of a group.  We use group
signature schemes to guarantee anonymity within a group, with the
authority being the group manager. 
From a group signature, only the group manager can learn the identity of the 
signer in the group. 
Note that the concept of groups, however, requires us to design an effective 
group division policy to optimally preserve users' location privacy. 
We discuss a solution in Sect.~\ref{Sec:discussion}.

Users collect their locations and anonymously send them to the toll
server, together with locations signed using a group signature scheme.
To prevent the authority from learning user locations
while opening signatures to resolve disputes, only the hashed values of locations are signed.
When it comes time to pay the toll, the server publishes the
hashed locations collected and the related fees. Fees are encrypted
with a homomorphic cryptosystem. Clients identify the fees that
correspond to their locations, add them up (homomorphically), and return the 
summation as their payments which, in turn, are opened by the server. 
%
If the summation of user payments does not match the summation of the fees of
collected locations, the server asks the authority to find out the dishonest
user. Together with the request, it sends to the authority the received 
group signatures, the corresponding encrypted fees,
and user payments. The authority cannot deduce locations from the
fees because of the homomorphic cryptosystem but, based on the
location signatures and the fees, it can compute (homomorphically)
the real toll of each user, and can thus identify the misbehaving
users: they are those whose committed payments differ from the real tolls.


We prove that our system is correct (users always pay their usage to the
server) and assures accountability (originators of misbehaviours can always be 
found and an effective punishing policy can be run to avoid repetitive 
misbehaviours). Furthermore, our system enforces
unlinkability between users and their locations. 

\vspace{2mm}
\noindent
{\bf Structure of the paper.}
Sect.~\ref{Sec:models} describes the participants of our system, 
the threat model and our assumptions, and it states the security 
goals of our design.  Sect.~\ref{Sec:GS} recalls group signature
schemes and other cryptographic primitives we adopt. Our ETP
system is fully described in Sect.~\ref{Sec:protocol}.
Sect.~\ref{Sec:analysis} defines the security properties and shows their 
enforcement by our system.
We conclude our paper in Sect.~\ref{Sec:discussion} with some discussions and 
ideas for future work.

\section{System Model}
\label{Sec:models}

\begin{figure}[!htp]
   \centering
   \includegraphics[scale=0.28]{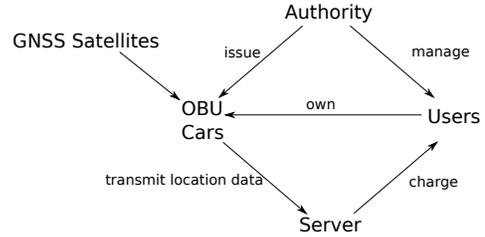}
   \caption{\label{Fig:structure}The relationship among the principals.}
\end{figure}

\vspace{2mm}
\noindent
{\bf Principals.} 
Our system consists of the following four principals: \emph{users},
\emph{their cars} with on board units (\emph{OBUs}), \emph{the authority}, and \emph{the toll 
server} (see Fig.~\ref{Fig:structure}).
       
Users own and drive cars, and they are responsible for toll payments.
To be entitled to use the electronic tolling service, a user brings
his car to the authority, which registers the car and installs an OBU
on it. The authority is a governmental department trusted by both
users and the toll server. It also builds up the group signature
scheme and manages groups of users.
An OBU computes the car's locations using GNSS satellites 
and stores them in a USB stick, which interfaces the OBU and contains
security information. It transmits location data to the toll server,
which is a logical organisation that can be run by multiple agents.
The server collects location data and computes the fee for each 
location record. 
It can also contact the authority to resolve disputes with insolvent users.

\vspace{2mm}
\noindent
{\bf Adversary model.} 
Considering the deployment environment 
of ETP systems, the possible threats can come from: 
1) manipulated OBUs, which generate false location tuples; 
2) dishonest users, who (partially) avoid to pay for their road usage; 
3) dishonest toll servers, who intend to increase their revenue and breach 
users' privacy; 
4) the honest but curious authority.

\vspace{2mm}
\noindent
{\bf Assumptions.} 
In our system, dishonest servers perform any actions so as to satisfy
their strong economic motivation and curiosity. They can deviate from
the protocol specification and collude with other attackers.  The
attackers considered in the system follow the Dolev-Yao intruder
model~\cite{DY83}. Specifically, they have full control over the
network, which means they can eavesdrop, block and inject messages
anywhere at anytime.  However, an encrypted message can never be
opened unless they have the right key. The identification based on the
message transmission is out of our scope in this paper. We assume that
location tuples are transmitted to toll servers anonymously. This can
be achieved by the architecture in~\cite{HGX06}, for instance, which
uses a communication service provider to separate authentication from
data collection.  The authority is supposed to be curious but not to
collude with any other participants.

It has been shown that users' moving traces can be reconstructed
from anonymised positions, e.g., by multi-target tracking
techniques or taking into account user mobility
profiles~\cite{HGXA07,STBH11}, and users' private information can thus be
inferred~\cite{K07}.  However, we observed from experiments in the
literature that tracking users remains difficult in practice,
especially when the intervals between transmissions are big (about
one minute) and the number of traveling users is not small. Therefore,
similar to VPriv, in this paper, we focus on privacy leakage from
ETP systems without considering the above mentioned techniques and
attacks.

\vspace{2mm}
\noindent
{\bf Security Properties.} 
In addition to the aim of reducing communication and computation overheads,
our system should employ proper measures to protect honest users and servers. 
Referring to what other systems
achieve (e.g.,~\cite{TDK07,PBB09,BRT10}), we address the following 
security properties:
\hfill \\[1ex]
{\it Correctness.} Clients pay for their own road usage and the server
collects the right amount of tolls.\hfill \\[1ex]
{\it Accountability.} If a malicious action that deviates from the
specification of the system occurs, sufficient evidence can be
gathered to identify its originator.\hfill \\[1ex]
{\it Unlinkability.} An intruder cannot link a given location record
to its generator.


\section{Cryptographic Primitives}
\label{Sec:GS}
\vspace{2mm}
\noindent
{\bf Group Signature Schemes.}
Group signatures~\cite{CV91} provide the signers anonymity among a group of users.
A group signature scheme involves group members and a group
manager. The task of the group manager is to organise the
group, to set up the group signature infrastructure and to reveal the
signer if needed. The signature of a message, signed by a group
member, can be verified by others based on the group public key while the
identity of the signer remains secret. 

Group signature schemes consist of at least the following five
functions: {\sc Setup}, {\sc Join}, {\sc Sign}, {\sc Verify} and {\sc Open}.
The function {\sc Setup} initialises the group public key,
the group manager's secret key and other related data. The procedure
{\sc Join} allows new members to join the group. Group members call
the function {\sc Sign} to generate a group signature based on their
secret keys.  The {\sc Verify} function makes use of the group public
key to check if the signature is signed by a group member.  The
function {\sc Open} determines the identity of the signer based on the
group manager secret key.

We take group signature schemes as an essential building block of our system 
because they have the following properties, which effectively meet our 
security goals. 
\begin{description*}
  \item [{\sc Correctness}] Signatures produced by a group member using {\sc 
  Sign} must be accepted by {\sc Verify}.
  \item [{\sc Unforgeability}] Only group members are able to sign messages on 
  behalf of the group.
  \item [{\sc Anonymity}] Given a valid signature of some message, identifying 
  the actual signer is unfeasible for everyone but the group manager.
  \item [{\sc Unlinkability}] Deciding whether two different valid signatures 
  were computed by the same group member is unfeasible.
  \item [{\sc Exculpability}] Neither a group member nor the group manager can 
  sign on behalf of other group members.
  \item [{\sc Traceability}] The group manager is always able to open a valid 
  signature and identify the actual signer.
  \item [{\sc Coalition-resistance}] A colluding subset of group members cannot 
  generate a valid signature that the group manager cannot link to one of the 
  colluding group members. 
\end{description*} 

There are some other properties we desire as well, e.g., efficiency and
dynamic group management. Efficiency contains the length of signatures
and computation time of each function, which determines the
feasibility of our system.  Dynamic group management enables users
to join and quit at any time when they are new or not satisfied with
current groups.

In the last decade, efficient group signature schemes with new fancy
features have been developed, e.g., group message authentication~\cite{PW10}
and group signcryption~\cite{ACHO11}. We will make use of an abstract
version of group signature schemes in the description of our system
as it can adopt any group signature scheme as long as it has the
required security properties.  Moreover, some of the schemes may improve the
security of our system.  For instance, an efficient group
signcryption scheme can prevent attackers from eavesdropping users'
location signatures over the network.

\vspace{2mm}
\noindent
{\bf Homomorphic Cryptosystem.}
Besides group signatures, we employ another cryptographic
primitive -- homomorphic public key cryptosystem.  Let $\ca{E}_X(m_1)$
and $\ca{E}_X(m_2)$ be two cipher texts of message $m_1$ and $m_2$
encrypted by agent $X$'s public key. Then we
have $\ca{E}_X(m_1)\cdot\ca{E}_X(m_2)=\ca{E}_X(m_1+m_2)$. There are
many cryptosystems with such properties, e.g., Paillier cryptosystem~\cite{P99}.

\vspace{2mm}
\noindent
{\bf Cryptographic Hash Function.}
In our system, we also use cryptographic hash functions,
which are publicly known and satisfy the minimum security requirements -- 
preimage resistance, second preimage resistance and 
collision resistance.  

\section{Our ETP System}
\label{Sec:protocol}   
We start with an informal description of our toll process, 
and proceed with notations and specifications of the protocols involved. 

\subsection{Overview} 
Our system is organised in four phases. The first phase is about the
service subscription and set-up. A user first signs a contract with
a toll server.  Both users and the server employ public key
encryption to secure their mutual communications.
When users contact the authority to join a group, the authority
assigns them to groups according to a group division policy
(see Sect.~\ref{Sec:discussion}).  Clients' private keys for group
signatures are also established during the phase.
After this, the server is informed of the groups containing its 
users and the corresponding group public keys.
  
The second phase is about collecting location data.  During driving,
OBUs compute their location and time which, together with the group
name, form what we call a \emph{location tuple}. OBUs periodically
send location tuples and the corresponding group signatures on the
hash values of the location tuples (called \emph{location signatures}) to the
server, who stores them in its location database.
  
The third phase is about calculating tolls. At the end of a toll
session, each user contacts the server by using a user interface
through browsers (not OBUs).  For any location tuple of a group, the
server publishes a \emph{fee tuple} consisting of the hashed location
tuple, and the corresponding cipher text of its fee using the chosen
homomorphic cyrptosystem.  A user selects his own fee tuples and
computes the cypher text of his toll payments by multiplying the
corresponding encrypted fees.  The server collects and decrypts all
users' toll payments.

The fourth phase is about resolving a dispute. 
This phase takes place \emph{only} when the 
sum of user payments \emph{in a group} is not equal to the sum of all location tuples' fees.
The authority is involved to determine the misbehaving user(s). 
The server sends the fee tuples and location signatures to the authority. 
When location signatures are opened, for each user, the authority 
identifies the encrypted fees belonging to his location tuples, 
whose multiplication is compared to the committed encrypted toll 
from the user.  Any inequality indicates that the corresponding user has 
cheated on his toll payment. 

\subsection{Notations}
Tab.~\ref{Tab:notation} summarises our notations.  With $c$, $S$, and
$A$ we indicate a user, the server, and the authority, respectively.
With $f(\ell, t)$ we indicate the fee to be paid when passing location
$\ell$ at time $t$, while $\client{\claimed{\cost}}$ is the amount of
fees that $c$ comitted to pay after the toll session $\mathit{sid}$.
$\mathit{Sig}_c(m)$ denotes the signature on message $m$ signed by
$c$, and $Gs_c(m)$ denotes the group signature of $c$ on message $m$.
For other cryptographic primitives, we use standard notations.

\begin{table}
\caption{\label{Tab:notation}Notations.} 
   \centering
   \begin{tabular}{r||p{5.8cm}}
     \hline  
     $f(\ell, t)$ & The fee of passing position $\ell$ at time $t$\\
     $\client{\claimed{\cost}}$ & The comitted toll payment of user $c$\\
     $sid$ & The identifier of the toll session\\
     $Sig_X(m)$ & Signature of message $m$ generated by a principal $X$ \\
     $Gs_c(m)$  &Group signature of message $m$ generated by a group member $c$\\
     $gpk(\mathcal{G})$ & The group public key of group $\mathcal{G}$\\
     $pk(X)$ & The public key of a principal $X$\\
     $sk(X)$ & The private key of a principal $X$\\
     $h(m)$& The \emph{hash} value of message $m$\\
     $\ca{E}_X(m)$ & The message $m$ encrypted with homomorphic 
     encryption with $X$'s public key\\
     $\mathit{Enc}_{pk(X)}(m)$ & The message $m$ encrypted with 
     $X$'s public key $pk(X)$\\
     \hline
 \end{tabular}
\end{table}

 \subsection{Protocol Specifications}
 \label{ssect:protocols}
Here we specify the four protocols that implement
the phases of our system, namely:
\emph{Set-up}, \emph{Driving}, \emph{Toll Calculation}, and
\emph{Dispute Solving}.
In the following discussion, we fix a group ${\cal G}$.

\vspace{2mm}
\noindent
{\bf Phase 1: Set-up.}
This protocol accomplishes two tasks. The first task is to establish
the public key infrastructure between the users, the server and the
authority.  The second task is to set up the group infrastructure.
The details are given in Appendix B.
  
\vspace{2mm}
\noindent
{\bf Phase 2: Driving.}
The driving protocol specifies how users periodically transmit
location tuples and location signatures to the server.  Let
$\tuple{\ell,t,\mathcal{G}}$ be a location tuple.  A message from
user $c$, a member of group $\mathcal{G}$, is denoted by
$(\tuple{\ell,t,\mathcal{G}}, {\it Gs}_{c}(h(\ell,t))$.  After
receiving this message, the server verifies ${\it Gs}_c(h(\ell,t))$
using the group public key $gpk(\mathcal{G})$.  If valid, 
the received message is stored.

\begin{figure}[h!]
   \centering
   \includegraphics[scale=0.33]{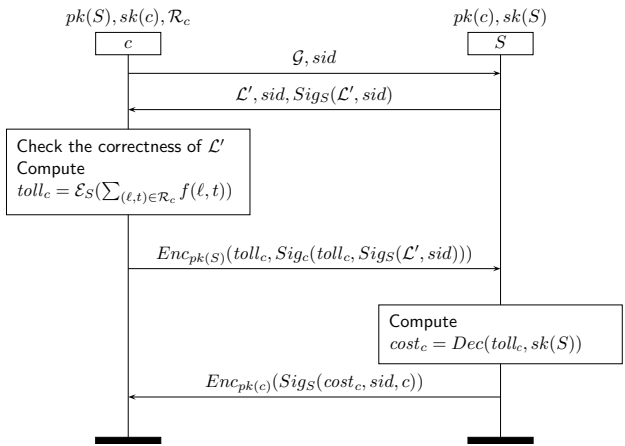}
   \caption{\label{Fig:deducing} The \emph{Toll Calculation} protocol.}
\end{figure}

\vspace{2mm}
\noindent
{\bf Phase 3: Toll Calculation.}  
This protocol aims to reach an agreement on toll payments between the
server and its users.
Let $\mathcal{L}'$ be the set of fee tuples of group $\mathcal{G}$.
An element in $\mathcal{L}'$ is of the form $(h(\ell, t),
\ca{E}_S(f(\ell, t)))$.  We use $\ca{R}_c$ to denote the locations
user $c$ has travelled and which are stored on the USB stick. We depict
the protocol in Fig.~\ref{Fig:deducing}.
  
In the server's response to $c$'s request, the server's signature on
$\mathcal{L}'$ is used to indicate that the fee tuples originate from the server. 
The user verifies the validity of this signature. By looking up the hash 
values of his locations, the user identifies his set of fee tuples. 
Then, for each of his location tuples $(\ell,t)$, the user computes
$\ca{E}_S(f(\ell, t))$ and compares it with the one in the
corresponding fee tuple.  If they are the same then the fee tuple is
correct.  If all his fee tuples are correct, he computes
$${\it toll}_c=\prod_{(\ell,t)\in\ca{R}_c}\ca{E}_S(f(\ell,t))$$
As the encryption is homomorphic, we have  
$${\it toll}_c=\ca{E}_S(\sum_{(\ell,t)\in\ca{R}_c}f(\ell,t))$$

The user then sends back to the server his signature on ${\it toll}_c$ and
$Sig_S(\ca{L}',sid)$. This signature indicates that
user $c$'s toll payment in toll session $sid$ is encrypted as ${\it toll}_c$
and computed based on $\mathcal{L}'$ issued by the server.
After receiving the user's response, the server verifies $c$'s
signature before sending back its signature of $c$'s toll payment
decrypted from ${\it toll}_c$.

Note that the set of fee tuples $\ca{L}'$ can be published in a repository in 
practice and accessed by authorised users. In this way, users are able
to download and check their validity without the need to be connected to the server.
Therefore, the real-time communication overhead is small. 

\begin{figure}[!htbp]
  \centering
  \includegraphics[scale=0.35]{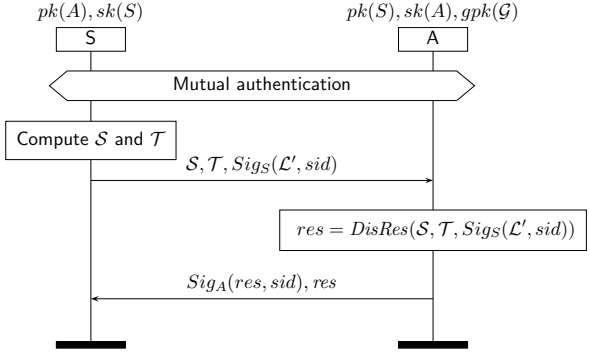}
  \caption{\label{Fig:DisSolving} The \emph{Dispute Resolving} protocol.}
\end{figure}

\vspace{2mm}
\noindent
{\bf Phase 4: Dispute Resolving.}  
In this protocol, with the help of the authority, the server finds the
cheating users and the amount of tolls unpaid. The server initiates
the dispute only when, with respect to a group, the sum of committed
payments is not equal to the sum of fees of all location tuples. Let
$\ca{L}$ be the set of location tuples of group $\ca{G}$.  This
condition can be formally described as
$$\sum_{c\in\ca{G}}\client{\claimed{\cost}}\neq\sum_{(\ell,
  t)\in\ca{L}} f(\ell,t)$$
A dispute will involve the authority, who can 
link a location signature to its signer. At the beginning of the
dispute resolution, the server constructs two sets $\ca{S}$ and $\ca{T}$.
$\ca{S}$ consists
of the hash values of the location tuples, the corresponding encrypted fees,
and the signatures of the location tuples that the server has received
in phase 2, that is:
\[
\ca{S} = 
\{
  \tuple{
  h(\ell, t),
  \hencr{\mathit{fee}(\ell,t)}{S}, 
  G_{{\ca{S}_{c}}}(h(\ell, t))}\mid
 \forall (\ell,t)\in{\cal L},\forall c\in{\cal G} 
\}
\]
$\ca{T}$ consists of the users' toll payment, that is:
\[
\ca{T} = \{  \tuple{c,{\it toll}_c, Sig_c({\it
    toll}_c,Sig_S(\ca{L}', sid))}\mid
\forall {c\in\ca{G}}\}
\]
We depict the protocol in Fig.~\ref{Fig:DisSolving}.
Note that for the sake of simplicity, we do not show the cryptographic details.

The critical part of the protocol is the function ${\it DisRes}$,
which is shown in detail in Alg.~\ref{Alg:dispute}. 
We use function ${\it checksign}(sign, m, pk)$ to check if the ${\it sign}$
is a signature of $m$ using $pk$ and group signature functions
{\sc Verify} and {\sc Open} work as described in Sect.~\ref{Sec:GS}.
The check on set $\ca{T}$ (line 5--8) and verification of location
signatures (line 10--11) exclude the possibility of modifying users'
toll payments by the malicious server. Each user's real toll payment
(i.e., $\client{\real{\toll}}$) is computed in lines 13--14. The
equivalence to the user's committed one (i.e.,
$\client{\claimed{\toll}}$) means the user pays the right amount,
otherwise, he is cheating.

After acquiring ${\it res}$ from the authority, the server can obtain
cheating users' real toll by decrypting $\client{\real{\toll}} =
\hencr{\client{\real{\cost}}}{S}$ and the amount of tolls unpaid as
well. Note that after resolving, the authority learns nothing about
users' locations except for the number of location records of each
user in that particular group.

\begin{algorithm}[!h]
  \caption{\label{Alg:dispute} Function ${\it DisRes}.$} 
  
  \begin{algorithmic}[1]
    \STATE {\bf Input}: $\ca{S},\ca{T},Sig_S(\ca{L}',sid)$ 
    \STATE {\bf Output}: $\mathcal{R}$
    \STATE ${\it res}:=\emptyset$;
    \STATE $\client{\real{\toll}}:= 0$;
    \FORALL {$(c,\toll, \mathit{sign})\in\ca{T}$}
       \STATE {\bf if} $\mathit{checksign}(\mathit{sign},(\toll, Sig_S(\ca{L}',sid)),pk(c))
              = {\it false}$ {\bf then}
       \STATE     ~~~~~~~~{\bf return} `check of $\ca{T}$ failed' ;
    \ENDFOR
    \FORALL  {$(\mathit{hashLoc, feeLoc, gsign})\in\ca{S}$}
       \STATE {\bf if} {\sc Verify}$(\mathit{gsign,hashLoc})={\it false}$ {\bf then} 
       \STATE     ~~~~~~~~{\bf return} `Faked location signatures' ;  
       \STATE {\bf else}  
       \STATE     ~~~~~~~~$c=${\sc Open}$(\mathit{gsign})$ ;
       \STATE     ~~~~~~~~$\client{\real{\toll}}$ 
                           := ({\bf if}~$\client{\real{\toll}} = 0$~{\bf then} 
                                       ~$\mathit{feeLoc}$~{\bf else}~
                                        $\client{\real{\toll}} \cdot \mathit{feeLoc}$);
    \ENDFOR
    \FORALL {$\client{\real{\toll}} \neq \client{\claimed{\toll}}$}
       \STATE ${\it res}:={\it res}\cup \{(c, \client{\real{\toll}})\}$ ;
    \ENDFOR
    \STATE {\bf return} $res$
  \end{algorithmic}
\end{algorithm}

\section{Security Properties \& Analysis}
\label{Sec:analysis}
In this section we define precisely what we mean by correctness, accountability
and unlinkability, and briefly discuss why our system
satisfies each of them. The full proof of our main theorem is given in Appendix A.

\vspace{2mm}
\noindent
{\bf Correctness.} 
Correctness means that the server can collect the right amount of
tolls and all users pay their tolls exactly. There are two
underlying assumptions according to practical toll scenarios. One is
that a user has no intention to pay more than his actual tolls while the
other is that the server wants no loss of users' tolls.

Let $\client{\real{\cost}}$ be the real amount of tolls that user $c$
should pay and ${\it pay}_c$ be the amount of tolls that user $c$ actually
pays to the server after phase 4 of our system.  Let
$\real{\toll}_{\ca{G}} = \sum_{c\in\ca{G}}{\client{\real{\cost}}}$ and
be the real amount of tolls from all users and ${\it pay}_\ca{G}
=\sum_{c\in\ca{G}}{\it pay}_c$ the amount of tolls that the server actually
collects from the group $\ca{G}$ after phase 4 of our system.
The property of correctness
can be defined as follows:
\begin{definition}[Correctness]
\label{def:correctness}
Suppose the server wants no loss of users' tolls and users have no
intention to pay more than their tolls, then for any $c\in\ca{G}$ it
holds that (${\it pay}_c=\client{\real{\cost}}$), and for the server it holds
that (${\it pay}_\ca{G}=\real{\cost}_\ca{G}$).
\end{definition}

In our system, whenever a user has paid less, the server initiates 
the dispute resolving protocol with the authority who would give the 
correct toll of that user. Meanwhile, because of the properties of group
signatures, e.g., {\sc Unforgeability} and {\sc Exculpability}, the server 
is unable to charge more locations than the ones users submitted.

\vspace{2mm}
\noindent
{\bf Accountability.} 
This property means that upon detection of malicious behaviour,
our system can identify which principal has misbehaved.

Let $\ca{B}$ be the set of all potential misbehaviours from the attackers in 
our system. So relation $\ca{A}=\ca{B}\times\ca{U}$ represents all possible 
attacks and the corresponding attackers. In our system, 
$\ca{U}=\ca{C}\cup\{S, A\}$. Let ${\it attacker}:\ca{A}\rightarrow\ca{U}$ be the 
function mapping an attack to the attacker, e.g., ${\it attacker}((\beta,c))=c$.
Let $E$ be the set of evidences during the run of our system and 
$\ca{P}(E)$ the power set of $E$. 
The definition of \emph{accountability} is given as follows:
\begin{definition}[Accountability]
Let $\ca{A}'\subseteq\ca{A}$ be the attacks that actually happen during the 
execution of our system in a toll session. For any $\alpha\in\ca{A}'$, our 
system is able to provide a set of evidences $E'\in\ca{P}(E)$ and there 
exists a function ${\it find}:\ca{P}(E)\times\ca{A}\rightarrow\ca{U}$ such 
that ${\it find}(E',\alpha)=attacker(\alpha)$.
\end{definition}

At steps where attackers may misbehave, our system provides sufficient
evidence to find the originators. For instance, when the server did not
send a user's toll payment to the authority on purpose, the server's 
signature on the user's payment could be taken as the evidence to prove the 
server's misbehaviour.

Despite the fact that our system assures accountability,
resolving disputes is still a costly step. We can establish a proper
punishment policy to discourage misbehaviours. This in turn also improves
the efficiency and performance of our system.  For instance, by
punishing the cheating users, the frequency of dispute resolving can
actually be very small in practice.

\vspace{2mm}
\noindent 
{\bf Unlinkability.}
Unlinkability holds when, from the information learned from the
execution of our system, the attacher cannot decide whether a user 
has travelled on any location. Proving unlinkability is equivalent to
prove that the adversary cannot distinguish the cases when two users 
swap their location information.

In order to enforce this property, we should consider two aspects. 
First, from all messages learned after the execution of our system, 
the attackers cannot link any location to its originator.
For malicious servers and users, the properties of group signature schemes,
i.e., {\sc Anonymity} and {\sc Unlinkability} guarantee this property. 
With regards to the honest but curious authority, implying that it
does not collude with
other attackers and only learns the hash values of locations, the property of 
the hash function enforces unlinkability.
Second, the communication process of the system does not give any information 
about linkability. This means, the attacker cannot break unlinkability
through analysing differences between executions of the system.
In order to check the satisfaction of unlinkability w.r.t.\ this situation, 
we apply the approach of formal verification (see Appendix A).

\smallskip
We now give the main theorem showing that our
ETP system satisfies the defined properties.
The full proof of the theorem is given in Appendix A.
\begin{theorem}
\label{thm}
Our ETP system guarantees correctness, accountability and unlinkability.
\end{theorem}

In Appendix C we also verified the relevant secrecy and authentication
properties.

\section{Discussion \& Conclusion}
\label{Sec:discussion}
In this paper, we have proposed a simple design for ETP systems which preserves 
users' anonymity within groups. 
In our system, our main design goal is to balance users' privacy with 
communication and computation overhead:
a large group means better privacy for the users,
while this gives rise to more overhead
when running the system.
With the help of group signature schemes and a homomorphic cryptosystem,
our system is proved to guarantee correctness, accountability and unlinkability.
To be complete, we still have the following issues to address.

\vspace{2mm}
\noindent
{\bf Comparison with VPriv.}
As mentioned in Sect.~\ref{Sec:intro}, our system resembles VPriv~\cite{PBB09}
that the server collects locations.  However, VPriv imposes a
relatively high burden to users and the server.

Compared with VPriv, in our system, the communication
overhead between users and the server is reduced.  Clients are
divided into groups which leads to a smaller set of fee tuples which
is returned from the server to users. The toll calculation protocol
also reduces the number and the size of messages during the interaction.
Second, we apply the principle of separation of duties in our system,
namely the authority takes the responsibility to find cheaters.
Hence, the server and users are released from a heavy computation overhead
by avoiding running zero-knowledge proof protocols as in VPriv.
Resolving disputes needs to open all location signatures, which is 
time consuming for the authority. However, with punishment policies and
the accountability property of  our system,
the authority can have a very low frequency of resolving disputes.

\vspace{2mm}
\noindent
{\bf Group management.}
A good group management policy can improve the protection of users' privacy 
in our system.
In principle, groups should be chosen to maximise the difficulty for the 
adversary to construct users' traces.
One way to achieve this goal is to group people according to `similarity' 
criteria based on multi-level hierarchical structure, as proposed in~\cite{GBW07}.
For instance, at the root level, we have the group of all users in a city. 
Subgroups at the next level contain those that usually travel in the
same region. At lower levels the subgroups can include users having a
similar driving style. Other factors can be considered as well,
e.g., driving periods, car models, etc.
The information needed to group people is collected by the
authority at the moment of registration. The provision of such information is
not compulsory but users are encouraged if they desire a better privacy protection. 

Dynamic group management, which enable users to change their group memberships,
is also necessary. For instance, users move to another city or they are not 
satisfied with their current group.  
To find the optimal group size which can protect users' 
anonymity is part of our future work.
Note that if a user has joined in multiple groups, the 
similarity between his travel records of these groups would decrease his anonymity. 

\vspace{2mm}
\noindent
{\bf Tamper resistant devices vs.\ spot checks.}
In order to ensure OBUs are not manipulated by users, e.g., to
transmit false locations, we have to consider possible solutions. One
way is to utilise devices that are tamper resistant. However, users can
always turn off the device. Therefore, as discussed in VPriv and
PrETP, we can use sporadic random spot checks that observe some
physical locations of users. A physical observation of a spot check
includes location, time and the car's plate number. Let
$\tuple{\ell,t,pn}$ be an observation of car $pn$ whose owner is
user $c\in\ca{G}$.  Then there should be at least one location
record $\tuple{\ell',t'}$ of group $\ca{G}$ such that $\mid
t,t'\mid<\epsilon/2$ and $\mid \ell,\ell'\mid<\gamma\cdot\mid
t,t'\mid$ where $\epsilon$ is the interval between two transmissions
and $\gamma$ is the maximum speed of vehicles.  If there are no such
location tuples, then the server can determine that user $c$
has misbehaved. Otherwise, the server could send the tuples with nearby
locations to the authority to check if there is one belonging to $c$.
According to~\cite{PBB09}, a small number of spot checks with a high
penalty would suffice.

\vspace{2mm}
\noindent
{\bf Future work.}
In future, we plan to develop a prototype of our
system and conduct an experimental evaluation to compare 
its efficiency with PrETP and VPriv.
A recently proposed ETP system Milo~\cite{MMCS11}
provides techniques based on blind identity based encryption
to strengthen spot checks in PrETP~\cite{BRT10} to
protect against large-scale driver collusion.
It is interesting to see how to adopt these techniques into our system.


%
\bibliographystyle{abbrv}
\bibliography{groupETP}  

\appendix
\section*{Appendix A: Proof of Theorem~\ref{thm}}

\begin{lem}
\label{lem}
Let $\ca{L}'$ be the set of fee tuples. 
If the server wants to collect no less tolls and users have no
intention to pay more than their tolls, then our system guarantees:
\begin{enumerate}
\item for any two sets of fee tuples $\ca{L}'_1$ and $\ca{L}'_2$ sent to  
users $c_1$ and $c_2$ from $\mathcal{G}$ in phase 3, 
$\ca{L}'_1=\ca{L}'_2$;
\item for each $(h(\ell, t), \ca{E}_S(\it{fee}))\in\ca{L}'$, 
${\it fee}= f(\ell,t)$;
\item $\ca{L}'$ consists of all location tuples sent by users.  
\end{enumerate}
\end{lem}
{\sc Proof.}
We prove the three sub-lemmas one by one.
\begin{enumerate}
\item Suppose $\ca{L}'_1\neq\ca{L}'_2$. 
The server initiates the dispute resolving protocol when some users paid less.
The authority perform the checks on signatures ${\it sign}_{c_1}$ and 
${\it sign}_{c_2}$ (line 6 -- 7 in Alg.~\ref{Alg:dispute}). 
The server wants to finish resolution in order not to lose tolls.
This means both of the two checks succeed, which implies  
$Sig_S(\ca{L}',sid)=Sig_S(\ca{L}'_1,sid)=Sig_S(\ca{L}'_2,sid)$.
So we can get that $\ca{L}'=\ca{L}'_1=\ca{L}'_2$. Contradiction.  

\item Suppose fee tuple $(h(\ell, t), \ca{E}_S({\it fee}))$ with 
${\it fee}\neq f(\ell,t)$, which belongs to user $c$.
In toll calculation, users check the correctness of $\ca{L}'$ before 
computing their payments. 
If the server want no loss of the tolls,
it has to ensure $\ca{L}'$ is correctly computed. Otherwise, the related
users would refuse to pay and take the server's signature on $\ca{L}'$ as 
the evidence of the server's misbehaviour.
This guarantees that for $\tuple{\ell, t}$ of $c$,
there exists exactly one fee tuple $(h(\ell,t),\ca{E}_S(f(\ell,t)))$, which 
contradicts the assumption.  
 
\item Suppose in $\ca{L}'$, the server removes a fee tuple 
$\varphi$ belonging to user $c$.
From the above two lemmas, user $c$ receives the same set of correct fee 
tuples without $\varphi$. 
Then $c$ would give a smaller toll payment.
According to our system, user $c$ would not be taken as a cheater and the 
server then has to take the loss.
This contradicts our assumption that the server wants no less tolls, 
therefore, the server would send the complete set of fee tuples to users.

\end{enumerate}
%
{\bf Proof of Theorem~\ref{thm}:}\\
We prove the three properties one after another.

\vspace{2mm}
\noindent
{\bf Correctness.}
We prove correctness from two aspects w.r.t.\ users and the server, 
respectively.
\begin{enumerate}

\item We start proving, by contradiction, that our system enforces
  that each user pays his real toll, that is
  $\forall_{c\in\ca{G}}~\client{\paid} = \client{\real{\cost}}$.

  Suppose that a user $c$ that has skipped some fees while
  homomorphically summing up his fees, which ends in
  $\client{\claimed{\cost}} < \client{\real{\cost}}$. As users,
  initially, pay exactly the cost they have summed and claimed to the
  server (which, we recall reveals them in clear the due), it follows
  that $\client{\claimed{\cost}} = \client{\paid} <
  \client{\real{\cost}}$.  It follows (Lemma~\ref{lem}) that
  $\sum_{c\in\ca{G}}\client{\claimed{\cost}}\neq \sum_{(\ell,
    t)\in\ca{L}} f(\ell,t)$: the server has ground to start the
  dispute resolving protocol.  The authority computes
  $\client{\real{\toll}}$ (\ie $\hencr{\client{\real{\cost}}}{S})$
  compares it with $\client{\claimed{\toll}}$ (\ie
  $\hencr{\client{\claimed{\cost}}}{S}$), discovers that they are not
  equal and returns it to the server with an accusation for $c$.  The
  server knows now the unpaid tolls $\client{\real{\cost}} -
  \client{\claimed{\cost}}$ and he can claim the missing due back. So,
  eventually, the amount paid by the user must be $\client{\paid} =
  \client{\claimed{\cost}} + (\client{\real{\cost}}
  -\client{\claimed{\cost}}) = \client{\real{\cost}}$.

\item We prove that
  $pay_\ca{G}=\sum_{c\in\ca{G}}\client{\real{\cost}}$.
  Straightforwardly follows from {1.} and from $\paid_{\ca{G}} =
  \sum_{c\in\ca{G}}{\client{\paid}}$.


\end{enumerate}

\noindent
{\bf Accountability.}
In our system, the set $\ca{B}$ consists of the following misbehaviours:
\begin{itemize*}
 \item $\beta_1$: dishonest users send smaller toll payment to the server;
 \item $\beta_2$: malicious users refuse to pay tolls;
 \item $\beta_3$: the server attaches wrong fees to location tuples;
 \item $\beta_4$: the server sends false location tuples to the authority;
 \item $\beta_5$: the server sends less toll payments to the authority;
\end{itemize*}

We prove accountability is secured against misbehaviours in $\ca{B}$.
\begin{figure*}[!htp]
  \centering
 \hspace{-2.5cm} \includegraphics[scale=0.33]{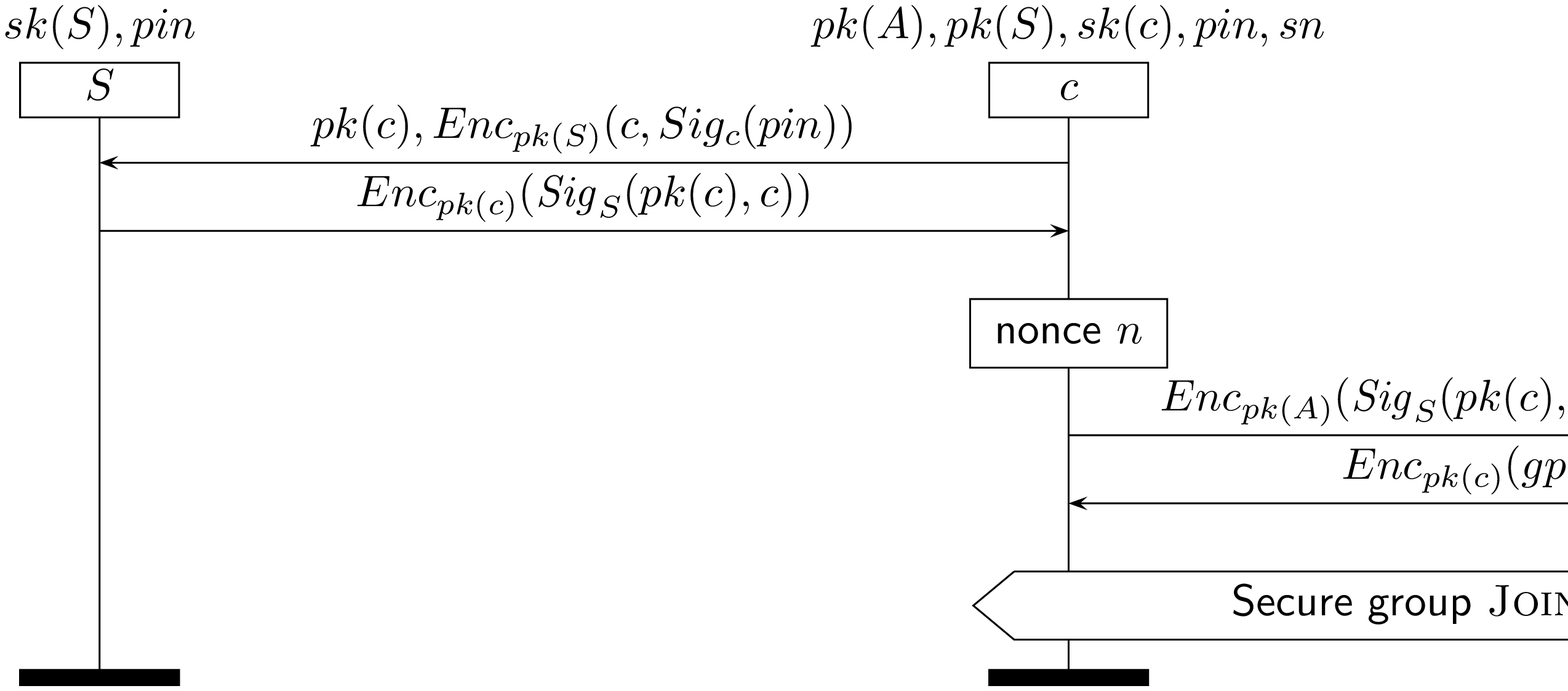}
  \caption{\label{Fig:setup} The \emph{Set-up} protocol.} 
\end{figure*}
\begin{enumerate}
\item Assume $\alpha=(\beta_1, c)$ happens, i.e.,
  $\client{\claimed{\cost}}<\client{\real{\cost}}$.  From the proof of
  Lemma 1, we know $c$ would be found out by the authority.  From the
  signed $res$ from the authority, the server learns
  $\client{\real{\toll}} = \hencr{\client{\real{\cost}}}{S}$.
  Together with user $c$'s signature on $\client{\claimed{\toll}}$ and 
  $\client{\real{\toll}} \neq \client{\claimed{\toll}}$, $c$ cannot deny his misbehaviour.

\item Assume $(\beta_2,c)$ happens.  We have the following two
  evidences -- (i) the server does not have user $c$'s signature on
  $\toll_c$; (ii) user $c$ cannot provide the server's signature on
  $\client{\claimed{\cost}}$ with $\ca{E}_S(\client{\claimed{\cost}})=\toll_c$.

\item Assume $(\beta_3, S)$ happens. 
In this attack, we have at least one fee tuple with a wrong fee. 
Let it be $(h(\ell,t),\ca{E}_S(fee))$ where $fee\neq f(\ell,t)$,
and the location $\tuple{\ell,t}$ belongs to user $c$.
When $c$ receives $\ca{L}'$, he identifies the corresponding fee to 
$\tuple{\ell,t}$ is $\ca{E}_S(fee)$ which is not equal to $\ca{E}_S(f(\ell,t))$.
Then he terminates the system. 
As the charging policy is public, with the server's signature on $\ca{L}'$,
the user can prove the server originates the attack. 

\item Assume $(\beta_4,S)$ happens. 
With the properties {\sc Unforgeability}, {\sc Exculpability} and 
{\sc Coalition-resistance} of group signature schemes, the server cannot
forge any location with a correct group signature from any honest user. 
Therefore, a failure of function {\sc Verify} (line 11 in 
Alg.~\ref{Alg:dispute}) suffices to determine the server's misbehaviour.

\item Assume $(\beta_5,S)$ happens.  Suppose the server omit user 
  $c$'s payments, i.e., $\toll_c$, which has been committed to the
  server during toll calculation.  In this case, the authority would
  return $res$ with $\client{\real{\toll}} =
  \ca{E}_S(\client{\real{\cost}})$.  However, $c$ has the server's
  signature on $\client{\claimed{\cost}}$ and (if indeed the user 
  has behaved correctly) he can prove that $\client{\claimed{\cost}} =
  \client{\real{\cost}}$.  This message is sufficient to prove the
  server's misbehaviour.
\end{enumerate}

\noindent
{\bf Unlinkability.}
For attacks on analysis of messages used during the execution of the 
system, the security of unlinkability is straightforward.
In our system, location tuples are hashed and signed through group signature 
schemes, which can be opened only by the authority. 
However, due to preimage resistance of hash primitives, the authority cannot 
learn users' locations through location signatures.
Meanwhile, the properties of {\sc anonymity} and {\sc unforgeability} ensure 
unlinkability against other attackers, i.e., the server and malicious users.

The second type of attacks on unlinkability is through observing the difference
between system executions where users' locations are varied.  
We start from defining unlinkability w.r.t.\ this type of attacks and proceed
proving our system's security using automatic formal verification.
We use processes to denote participants' behaviours in protocols. 
Let $C\langle\varphi\rangle$ be the process representing user $c$ originating
location record $\varphi$ and let $A$ be the process of the authority.
We use $C\langle\varphi\rangle\mid A$ to represent the parallel composition of 
these two processes, which admits all possible communications and interleavings.
The intuition behind unlinkability is that if any two users swap
a pair of locations, the adversary cannot observe the difference. 
\emph{Observational equivalence}, which defines indistinguishability between
two processes~\cite{AF01}, gives us an effective way to formalize unlinkability 
in our system.
Similar to the case of electronic voting~\cite{DKR09}, we need at least
two traveling users. Otherwise, an intruder can easily link all location 
tuples to one user.
\vspace{-1mm}  
\begin{definition}[Unlinkability]
 Assume that $\mathcal{G}$ is a group with at least two users $c$ and $c'$ and
 assume any two location records $\varphi$ and $\varphi'$. 
 Unlinkability between a location tuple and its generator holds if
 $$A \mid C\langle\varphi\rangle \mid C'\langle\varphi'\rangle\approx 
  A \mid C\langle\varphi'\rangle \mid C'\langle\varphi\rangle$$
\end{definition}

ProVerif~\cite{B01} is an efficient and popular tool for verifying security 
properties in cryptographic protocols. It takes a protocol modelled as a 
process in the applied $\pi$ calculus~\cite{AF01} as input and
checks whether 
the protocol satisfies a given property. Observational equivalence can 
be modelled and verified by ProVerif. Thus, the definition of 
unlinkability leads us to an automatic verification.
We have modelled our system and the unlinkability property, and got a positive 
result from ProVerif. This means our system preserves unlinkability. 
(ProVerif codes are available on request.)

\section*{Appendix B: The Set-up Phase}
We make use of two secrets to achieve the security goal of this phase -- 
\emph{the pin codes} and \emph{the serial numbers}. The former is generated
by the server for users to prove their legal access to toll service,
while a serial number is issued with each OBU as a secret between the authority
and a user. 
We take user $c$ as an example. 
Let $pin$ be his pin code and $sn$ the serial number of his OBU.
The Setup protocol is depicted in Fig.~\ref{Fig:setup}.
Upon receiving the user's public key, the server checks $c$'s signature on 
$pin$.  If valid, the server replies with its signature on the key, which the 
user sends to the authority subsequently when joining a group.
A replay attack on $c$'s message to the authority is not feasible, as the same 
group would be returned if the same request message arrives again.
Fig.~\ref{Fig:setup} does not include the last step where the server learns 
from the authority its users' groups and the group public keys,
since such information can be made public.

\section*{Appendix C: Secrecy and Authentication}
\label{Sec:verification}
We use ProVerif~\cite{B01} to formally prove that our system 
as a whole does not suffer from attacks on security and authentication.
The results are listed in Tab.~\ref{Tab:auth-sec}. 
(ProVerif codes are available on request.)

\vspace{-2mm}
\begin{table}[!htp]
\caption{\label{Tab:auth-sec}Verification of authentication and 
secrecy.}
  \centering
  \begin{tabular}{|l||c|c||c|}
    \hline
    \multirow{2}{*}{~~Protocols~~} & \multicolumn{2}{c||}{Authentication} & 
    \multirow{2}{*}{Secrecy}\\
    \cline{2-3} 
     & injective & non-injective & \\ 
    \hline
     setupCS & -- & $c$ \& $S$ & $pin$ \\ \hline
     setupCA & $A$ & $c$ & ~$sn$~\\ \hline
     Toll Calculation& --& $c$ \& $S$ & $\toll_c$\\ \hline
     Dispute resolving & $S$ \& $A$ & --& $\ca{T},res$ \\ \hline
 \end{tabular}  
\end{table} 
We use setupCS to denote the protocol between the server ($S$) and a user 
($c$) in the setup phase and setupCA is between a user and the authority ($A$).
We say a term is secret if the attacker cannot get it by eavesdropping and 
sending messages, and performing computations~\cite{B01}. 
For authentication, we consider two notions, namely, \emph{agreement} and the 
slightly stronger notion \emph{injective agreement}~\cite{L97}.
Agreement roughly guarantees to an agent $A$ that his communication
partner $B$ has run the protocol as expected and that $A$ and $B$
agreed on all exchanged data values. Injectivity further requires that
each run of $A$ corresponds to a unique run of $B$.

\end{document}